\documentclass[preprint,aip,jcp,amsmath,amssymb]{revtex4-1}
\usepackage[utf8]{inputenc}
\usepackage{amsmath}
\usepackage{graphicx}   
\usepackage{dcolumn}    
\usepackage{bm}         
\usepackage{hyperref}
\usepackage{color}
\usepackage{array}

\newcommand{\jcp}{J.\ Chem.\ Phys.\ }

\newcommand{\jpc}{J.\ Phys.\ Chem.\ }

\newcommand{\prl}{Phys.\ Rev.\ Lett.\ }

\newcommand{\jpcl}{J.\ Phys.\ Chem.\ Lett.\ }

\newcommand{\eqn}[1]{Eq.~(\ref{#1})}

\newcommand{\eqnn}[2]{Eqs.~(\ref{#1}) and (\ref{#2})}

\newcommand{\bra}[1]{\langle #1 |}
\newcommand{\ket}[1]{| #1 \rangle}

\begin{document}

\title{An alternative derivation of ring-polymer molecular dynamics transition-state theory} 
\author{Timothy J.~H.~Hele\footnote{Current address: Department of Chemistry and Chemical Biology, Cornell University, Ithaca, New York 14853, USA.}}
\affiliation{Department of Chemistry, University of Cambridge, Lensfield Road, Cambridge, CB2 1EW, UK.}
\author{Stuart C.~Althorpe\footnote{Corresponding author: sca10@cam.ac.uk}}
\affiliation{Department of Chemistry, University of Cambridge, Lensfield Road, Cambridge, CB2 1EW, UK.}
\date{\today}

\begin{abstract}
In a previous article [J.~Chem.~Phys.\ {\bf 138},  084108 (2013)], we showed that the $t\to 0_+$ limit of ring-polymer molecular dynamics (RPMD) rate-theory is also the $t\to 0_+$ limit of a new type of quantum flux-side time-correlation function, in which the dividing surfaces are invariant to imaginary-time translation; in other words, that RPMD transition-state theory (RPMD-TST) is a $t\to 0_+$ quantum transition-state theory (QTST). Recently, Jang and Voth [J.~Chem.~Phys.\ {\bf 144}, 084110 (2016)] rederived this quantum $t\to 0_+$ limit, and claimed that it gives instead the centroid-density approximation. Here we show that the $t\to 0_+$ limit derived by Jang and Voth is in fact  RPMD-TST.
\end{abstract}

\maketitle 

\section{Introduction}

Ring-polymer molecular dynamics (RPMD) rate-theory is a powerful method for calculating approximate thermal quantum reaction rates. The method has been applied to a variety of reactions, in both the gas and condensed phases,\cite{rpmd1,rpmd2,rpmd3,rpmd4,rpmd5,rpmd6,rpmd7,rpmd8,rpmd9,rpmd10,rpmd11,rpmd12} where it has been found to give a good approximation to the exact quantum result (where this is available) across a wide temperature range, from the classical to the deep-tunnelling regime.

The success of RPMD rate-theory was initially a mystery, as the method was proposed on a heuristic basis,\cite {rpmd1,rpmd2} and it was not clear how a method that involves classical molecular dynamics in an extended ring-polymer space could reproduce deep-tunnelling rates. A subsequent analysis\cite{jeremy} at low temperatures showed that the $t\to 0_+$ limit of the RPMD flux-side time-correlation function, i.e.\ the RPMD transition-state-theory rate (RPMD-TST), contains a quantum-Boltzmann ensemble of Feynman paths which fluctuate around the instanton\cite{inst1} (periodic orbit); this holds even for highly asymmetric reaction barriers, for which the earlier centroid-density approximation\cite{gillan,vcm} (which is the special case of RPMD-TST with a centroid dividing surface) breaks down.\cite{jeremy,vothas}

More recently, it was found that the RPMD-TST rate also emerges naturally as a  quantum transition-state-theory (QTST), corresponding to the $t\to 0_+$ limit of a new type of quantum flux-side time-correlation
 function.\cite{tim1,tim2,tim3} By placing the flux and side dividing surface in the same place in path-integral space, this function gives a non-zero $t\to 0_+$ limit, and by making these  surfaces invariant to imaginary-time translation, it gives the correct quantum Boltzmann statistics (thereby avoiding the problem of negative rates, encountered in the related classical Wigner expression\cite{wiggy}). It was further shown\cite{tim2} that this $t\to 0_+$ limit (i.e.\ RPMD-TST) gives the exact quantum rate in the absence of recrossing of the dividing surface (and of surfaces orthogonal to it in path-integral space), and gives an approximate upper bound to the exact quantum rate, which becomes an exact upper bound in the high-temperature limit (where classical TST is recovered as a special limiting case).

A recent paper by Jang and Voth\cite{jv} appears to contradict these findings; these authors derive the $t\to 0_+$ limit of the same quantum time-correlation function as in ref.~\onlinecite{tim1}, but claim to find that it gives the centroid-density approximation. Here we show that there is no such contradiction, because the $t\to 0_+$ limit obtained by Jang and Voth is in fact RPMD-TST. The article is structured as follows: Sec.~II summarises the key equations of RPMD rate theory and gives the  quantum time-correlation function of ref.~\onlinecite{tim1}; Sec.~III presents an analysis of the $t\to 0_+$ limit derived by Jang and Voth; Sec.~IV  concludes the article.

\section{Summary of previous results}

Here we summarise previous results from RPMD rate-theory and give the quantum time-correlation function introduced
in ref.~\onlinecite{tim1}, of which the RPMD-TST rate is the $t\to 0_+$ limit. We will confine the analysis to a one-dimensional system with classical Hamiltonian
\begin{align}\label{1d}
H={p^2\over 2m}+V(q)
\end{align}
It is straightforward to generalize these approaches to multi-dimensional systems.\cite{rpmd1,rpmd2,rpmd3,tim1,tim2,tim3}

\subsection{RPMD-TST}

For the system of \eqn{1d}, the RPMD Hamiltonian is 
\begin{align}
H_N=\sum_{i=1}^N{p_i^2\over 2 m}+U_N({\bf q})
\end{align}
in which ${\bf q}=\{q_1,\dots,q_N\}$ are a set of $N$ replicas of the system coordinate $q$, ${\bf p}=\{p_1,\dots,p_N\}$ are the conjugate momenta, and $U_N({\bf q})$ is the ring-polymer potential
\begin{align}
U_N({\bf q})=\sum_{i=1}^N{m(q_{i+1}-q_i)^2\over 2(\beta_N\hbar)^2}+V(q_i)
\end{align}
with $q_{i\pm N}=q_i$.
Clearly $U_N({\bf q})$ is the exponent in the standard path-integral expression\cite{chandler,parrinello,ceperley} for the quantum Boltzmann operator 
$\exp({-\beta \hat H})$. The dynamics generated by $H_N$ is fictitious, but satisfies two important criteria: it is exact in the limit $t\to 0$, and it preserves the quantum Boltzmann distribution. These properties allow one to apply (standard) classical rate theory in the extended phase space $({\bf p},{\bf q})$, to compute a rate coefficient which gives a lower-bound estimate of the $t\to 0_+$ flux through some
 {\em dividing surface} $f({\bf q})$;
   this initial flux is the RPMD-TST approximation to the quantum rate coefficient:
\begin{align}\label{rpmd1}
k^\ddag_{\rm RP}(T)Q(T)=\lim_{N\to\infty}{1\over (2\pi\hbar)^N}\int\!d{\bf p}\int\!d{\bf q}\,e^{-\beta_NH_N}\delta[f({\bf q})]{\dot f}({\bf q})h[{\dot f}({\bf q})]
\end{align}
where $Q(T)$ is the reactant partition function, and
\begin{align}
{\dot f}({\bf q})=\sum_{i=1}^N{\partial f({\bf q})\over \partial q_i}{p_i\over m}
\end{align}
is the $t\to 0_+$ flux through $f({\bf q})$ (and $h(x)$ denotes the Heaviside step-function, and  we use the notation $\int\!d{\bf q}=\int_{-\infty}^\infty dq_1\dots\int_{-\infty}^\infty dq_N$  throughout).

An important property of $f({\bf q})$ is that, in order to maximise the free energy, it must be invariant under cyclic permutation of the replicas, i.e. 
\begin{align}\label{perm}
{\cal P}_{i\rightarrow i+k}\,f({\bf q})=f({\bf q})
\end{align}
where ${\cal P}_{i\rightarrow i+k}$ indicates that each $q_i$ is moved to the position previously occupied by $q_{i+k}$.
A common choice of $f({\bf q})$ satisfying this condition is $f({\bf q})={Q}_0-q^\ddag$, where ${Q}_0=\sum_i^Nq_i/N$ is the ring-polymer {\em centroid} (centre of mass). This important special case of RPMD-TST is often referred to as the centroid-density approximation.\cite{gillan,vcm} As mentioned in the Introduction, the centroid dividing-surface works well above the cross-over temperature to deep tunnelling, but more general forms of $f({\bf q})$ need to be used at lower temperatures if the barrier is asymmetric (in which case the optimal dividing surface involves  ring-polymer stretch modes).\cite{jeremy} In the limit $N\to\infty$, \eqn{perm} is equivalent to making $f({\bf q})$  invariant to imaginary-time translation, provided $f({\bf q})$ is also a smooth function of imaginary time (see the Appendix), which we will assume in what follows.

Equation (\ref{rpmd1}) can be obtained in more compact form by integrating out the momenta ${\bf p}$, to give
\begin{align}\label{comp}
k^\ddag_{\rm RP}(T)Q(T)=\lim_{N\to\infty}{1\over 2\pi\hbar\beta_N}\left(m\over 2\pi\beta_N\hbar^2\right)^{(N-1)/2}\int\!d{\bf q}\,e^{-\beta_NU_N}\sqrt{B_N({\bf q})}\delta[f({\bf q})]
\end{align}
where
\begin{align}\label{bn}
B_N({\bf q})=\sum_{i=1}^N\left[{\partial f({\bf q})\over \partial q_i}\right]^2
\end{align}
normalises the flux. This expression will turn out to be useful in Sec.~III. 

\subsection{Quantum $t\to 0_+$ TST}
In ref.~\onlinecite{tim1}, we found a quantum flux-side time-correlation function whose  $t\to 0_+$ limit gives $k^\ddag_{\rm RP}(T)$. The standard forms of flux-side time-correlation function (obtained from linear response\cite{yama} or scattering theory\cite{mst}) give zero as $t\to 0_+$. This property was shown in ref.~\onlinecite{tim1} to be the result of putting the flux and side dividing surfaces in different locations in path integral space, with the result that the flux and side are initially decorrelated and therefore zero. When the flux and side dividing surfaces are in the same place, and when they are taken to be a smooth permutationally invariant function $f({\bf q})$ as defined above, then the resulting quantum flux-side time-correlation function $C_{\rm fs}(T,t)$ satisfies\cite{tim1}
 \begin{align}
k^\ddag_{\rm RP}(T)Q(T)=\lim_{t\to 0_+}C_{\rm fs}(T,t)
\end{align}
The simplest way to write out $C_{\rm fs}(T,t)$ is as the derivative of the corresponding side-side function
\begin{align}\label{qfs}
C_{\rm fs}(T,t)=-{d C_{\rm ss}(T,t)\over d t}
\end{align}
where
\begin{align}\label{quant}
C_{\rm ss}^{[N]}(T,t)=\lim_{N\to\infty}&\int\!d{\bf q}\int\!d{\bf \Delta}\int\!d{\bf z}\,h[f({\bf q})]h[f({\bf z})]\rho_N({\bf q},{\bf \Delta})\nonumber\\
&\times\bra{q_{i}-\Delta_{i}/2}e^{i{\hat H}t/\hbar}\ket{z_i}\bra{z_i}e^{-i{\hat H}t/\hbar}\ket{q_{i}+\Delta_{i}/2}
\end{align}
with
\begin{align}
\rho_N({\bf q},{\bf \Delta})=\prod_{i=1}^N\bra{q_{i-1}-\Delta_{i-1}/2}e^{-\beta_N{\hat H}}\ket{q_{i}+\Delta_{i}/2}
\end{align}
and 
\begin{align}
{\hat H}={\hat p^2\over 2m}+V({\hat q})
\end{align}
The $t\to\infty$ limit of $C_{\rm fs}(T,t)$ [of \eqn{qfs}] does not give the exact quantum rate, since one must also account for recrossing of 
dividing surfaces orthogonal to $f({\bf q})$ in path-integral space. However, it was shown in ref.~\onlinecite{tim2} that the flux through these orthogonal dividing surfaces is zero in the limit $t\to 0_+$, and thus that $k^\ddag_{\rm RP}(T)$ gives the instantaneous thermal quantum
 flux from reactants to products.

\section{The alternative derivation}

In ref.~\onlinecite{jv}, Jang and Voth rederived the $t\to 0_+$ limit of $C_{\rm fs}(T,t)$ and found that it gives\cite{suppress}
\begin{align}
k^\ddag_{\rm JV}(T)Q(T)=\lim_{t\to 0_+}C_{\rm fs}(T,t)
\end{align}
where
\begin{align}\label{jv}
k^\ddag_{\rm JV}(T)Q(T)=\lim_{N\to\infty}\,&{1\over 2\pi\hbar\beta_N}\int\!d{\bf q}\int\!d \eta\,{\tilde \rho}_N({\bf q},\eta)\delta[f({\bf q})]\nonumber\\
&\times\sum_{k=1}^N{\partial f({\bf q})\over \partial q_k}{T_{k-1}+2T_k+T_{k+1}\over 4}
\end{align}
with
\begin{align}
{\tilde \rho}_N({\bf q},\eta)=\prod_{i=1}^N\bra{q_{i-1}-T_{i-1}\eta/2}e^{-\beta_N{\hat H}}\ket{q_{i}+T_{i}\eta/2}
\end{align}
and
\begin{align}\label{tt}
T_i({\bf q})={1\over \sqrt{B_N({\bf q})}}{\partial f({\bf q})\over \partial q_i}
\end{align}
 After analysing \eqn{jv}, Jang and Voth concluded that it gave the centroid-density rate instead of RPMD-TST.

We now show that \eqn{jv} does in fact give RPMD-TST, i.e.\ that
\begin{align}\label{eq}
k^\ddag_{\rm JV}(T)\equiv k^\ddag_{\rm RP}(T)
\end{align}
We first note that Jang and Voth's analysis\cite{jv} considered only the special case of a centroid dividing surface. We therefore need to generalize the analysis to a smooth, permutationally invariant, $f({\bf q})$, which satisfies \eqn{perm} (which includes the centroid dividing surface as a special case). Exploiting first the smoothness of $f({\bf q})$, we note that the last term in \eqn{jv} can be replaced by $T_k({\bf q})$ (see the Appendix), such that \eqn{jv} simplifies to
\begin{align}\label{re}
k^\ddag_{\rm JV}(T)Q(T)=\lim_{N\to\infty}\,&{1\over 2\pi\hbar\beta_N}\int\!d{\bf q}\int\!d \eta\,{\tilde \rho}_N({\bf q},\eta)\delta[f({\bf q})]\sqrt{B_N({\bf q})}
\end{align}
[where we have used \eqnn{bn}{tt} to replace the sums over $T_k$ by $\sqrt{B_N({\bf q})}$]. A similar procedure allows us to evaluate 
${\tilde \rho}_N({\bf q},\eta)$ explicitly in terms of matrix elements over the position coordinates, replacing instances of $T_{k+1}+T_k$ by $2T_k$, to give
\begin{align}
{\tilde \rho}_N({\bf q},\eta)=&\left(m\over 2\pi\beta_N\hbar^2\right)^{N/2}\exp\left[-\sum_{i=1}^Nm(q_{i+1}-q_i)^2/2\beta_N\hbar^2\right]\nonumber\\
&\times e^{-\beta\Phi_N({\bf q})}e^{-\eta^2m/2\beta_N\hbar^2}e^{-g_N({\bf q})\eta/\hbar}+{\cal O}(N^{-1})
\end{align}
with
\begin{align}
\Phi_N({\bf q})=&\frac{1}{2N}\sum_{i=1}^NV(q_i+T_i\eta/2)+V(q_i-T_i\eta/2)\\
=&\left[{1\over N}\sum_{i=1}^NV(q_i)\right]+{\cal O}(\eta^2N^{-1})\label{vv}
\end{align}
(where the last line uses the property $T_i\sim N^{-1/2}$)
and 
\begin{align}
g_N({\bf q})=&{m\over 2\beta_N\hbar}\sum_{i=1}^N(q_{i+1}-q_{i})T_i({\bf q})
\end{align}
Equation (\ref{vv}) ensures that $V$ depends only on ${\bf q}$ in the limit $N\to\infty$, allowing us to integrate over $\eta$. Because of the cross term $g_N({\bf q})$, this integral will, in the case of a completely general (i.e.\ non-permutationally invariant) dividing surface,
give a complicated expression involving repulsive `springs' between all pairs of `beads' $q_i$. 
However, for the smooth, permutationally invariant, $f({\bf q})$, it is sufficient to note 
that
\begin{align}
\left({\cal P}_{i\to i+1}-1\right)f({\bf q})=\sum_{i=1}^N(q_{i+1}-q_{i}){\partial f({\bf q})\over \partial q_i}+{\cal O}[(q_{i+1}-q_{i})^3]
\end{align}
and that $q_{i+1}-q_{i}\sim N^{-1/2}$, from which it follows that the cross-term $g_N({\bf q})$ disappears in the limit $N\to\infty$.
 The integral over $\eta$ in \eqn{re} then closes up the matrix elements in ${\tilde \rho}_N({\bf q},\eta)$
  into an ensemble of intact ring-polymers, giving
\begin{align}
\int\!d \eta\,{\tilde \rho}_N({\bf q},\eta)=\left(m\over 2\pi\beta_N\hbar^2\right)^{(N-1)/2}e^{-\beta_NU_N({\bf q})}+{\cal O}(N^{-1})
\end{align}
Substituting this expression back into \eqn{re} gives the righthand side of \eqn{comp}, thus proving \eqn{eq}.\cite{error}

\section{Summary}

We have shown that ref.~\onlinecite{jv} gives an alternative derivation of RPMD-TST. There is thus no contradiction between the $t\to 0_+$ limits
 derived in refs.~\onlinecite{tim1} and \onlinecite{jv}, and we can be clear that RPMD-TST is a quantum transition-state theory (QTST), obtained
 as the $t\to 0_+$ limit of a quantum time-correlation function describing the flux through a dividing surface that is invariant to imaginary-time translation. As discussed in ref.~\onlinecite{tim2}, this does not imply that  RPMD-TST is a good approximation to all quantum reaction rates: RPMD-TST works if the reaction is direct, and if the temperature is not too far below the instanton cross-over temperature. There are of course many reactions for which these conditions apply, and the range of applications of RPMD rate-theory is constantly growing.\cite{rpmd1,rpmd2,rpmd3,rpmd4,rpmd5,rpmd6,rpmd7,rpmd8,rpmd9,rpmd10,rpmd11,rpmd12}

\begin{acknowledgments}
We acknowledge funding from the UK Science and Engineering Research Council. TJHH also acknowledges a Research Fellowship from Jesus College, Cambridge.
\end{acknowledgments}

\appendix

  \section*{Appendix: Smooth dividing surfaces}
  
 One can construct an $f({\bf q})$ which is a smooth function of imaginary time by making it depend on a finite set of free ring-polymer normal modes,      \begin{align}
{Q}_n&=\sum_{l=1}^NU_{ln}q_l, \quad n=0,\pm 1,\dots,\pm (M-1)/2
\end{align}
with
\begin{align}
 U_{ln} = N^{-1}\times
 \left\{
 \begin{array}{ll}
  1 & n=0 \\
  \sqrt{2} \sin(2\pi ln/N) & n=1,\dots,(M-1)/2 \\
  \sqrt{2} \cos(2\pi ln/N) & n=-1,\dots,-(M-1)/2
 \end{array}
 \right.\label{ttt}
\end{align}
(and note that we have normalised the modes such that $Q_0$ corresponds to the centroid). Taking $M=0$ makes $f({\bf q})$ a function of just the centroid; taking $M>0$ gives a more general dividing surfaces, such as is needed for asymmetric barriers below the cross-over temperature.\cite{jeremy}

  The smoothness of $f({\bf q})$ imposes relations between derivatives ${\partial f({\bf q}) / \partial q_i}$ and thus between
 $T_i({\bf q})$ for different values of $i$. From \eqn{tt}, it follows that
\begin{align}
T_i({\bf q})={1\over\sqrt{B_N({\bf q})}}\sum_{n=-(M-1)/2}^{(M-1)/2}U_{in}{\partial f({\bf Q})\over\partial Q_n}
 \end{align}
Substituting for $U_{in}$,  using trigonometric identities, and taking the limit $N\to\infty$ (whilst noting that $M$ is finite), we obtain
 \begin{align}
T_{i+1}({\bf q})=T_{i}({\bf q})+{\cal O}(N^{-1})
 \end{align} 
 This allows us to replace $(T_{i+1}+2T_i+T_{i-1})/4$ and $(T_{i+1}+T_i)/2$ by $T_i$ in Sec.~III.


\begin{thebibliography}{10}
\bibitem{suppress} Note that we have modified slightly the notation of ref.~\onlinecite{jv};
  also, we will usually omit writing the ${\bf q}$-dependence of $T_i({\bf q})$ to save space.
\bibitem{error} This result also shows that no error was made in ref.~\onlinecite{tim1} in evaluating the matrix elements over $\hat p$
  [which is one of the reasons suggested in ref.~\onlinecite{jv} for the supposed
   disagreement between $k^\ddag_{\rm RP}(T)$ and $k^\ddag_{\rm JV}(T)$].
\end{thebibliography}
\end{document}